\documentclass[12pt,preprint]{aastex}

\begin{document}

\title{An excursion set model of the cosmic web:  
       The abundance of sheets, filaments and halos}

\author{Jiajian Shen}
\affil{Department of Astronomy and Astrophysics, 
       The Pennsylvania State University, University Park, PA 16802}
\author{Tom Abel}
\affil{Kavli Institute for Particle Astrophysics and Cosmology, 
       Stanford Linear Accelerator Center, 
       2575 Sand Hill Road, MS 29, Menlo Park, CA 94025}
\author{H.J. Mo}
\affil{Department of Astronomy, University of Massachusetts, 
       Amherst MA 01003}
\and
\author{Ravi K. Sheth}
\affil{Department of Physics and Astronomy, 
       University of Pennsylvania, 209 South 33rd Street, 
       Philadelphia, PA 19104}

\begin{abstract}
We discuss an analytic approach for modeling structure formation in 
sheets, filaments and knots.  This is accomplished by combining 
models of triaxial collapse with the excursion set approach:  
sheets are defined as objects which have collapsed along only one 
axis, filaments have collapsed along two axes, and halos are objects 
in which triaxial collapse is complete.  
In the simplest version of this approach, which we develop here, 
large scale structure shows a clear hierarchy of morphologies:  
the mass in large-scale sheets is partitioned up among lower mass 
filaments, which themselves are made-up of still lower mass halos.  
Our approach provides analytic estimates of the mass fraction in 
sheets, filaments and halos, and its evolution, for any background 
cosmological model and any initial fluctuation spectrum.  In the 
currently popular $\Lambda$CDM model, our analysis suggests that 
more than 99\% of the cosmic mass is in sheets, and 72\% in 
filaments, with mass larger than $10^{10} M_{\odot}$ 
at the present time.  For halos, this number is only $46\%$.  
Our approach also provides analytic estimates of how halo abundances 
at any given time correlate with the morphology of the surrounding 
large-scale structure, and how halo evolution correlates with 
the morphology of large scale structure.  
\end{abstract}

\section{Introduction}
Recent observations and high resolution numerical simulations of 
structure formation show that, on large scales, the Universe 
is best thought of as a Cosmic Web \citep{BKP96}:  large scale 
sheets are traversed by filaments, which themselves intersect at 
knots \citep{BL88, BS00, BBS03, Sheth03, BB04, Shandarin04, BB05, CKC05}, 
and there are vast relatively empty voids in between 
\citep{GT82, KOSS83, MAM03, VHRD04}.  
The knots where filaments intersect are the locations of rich 
galaxy clusters, and these are often identified with the massive 
virialized dark matter halos found in simulations.  
Halos are the best-studied features of the cosmic web; 
they tend to be about two hundred times denser than the background 
universe, so they account for a small fraction of the volume of the 
Universe.  Simple analytical models have been very successful in 
understanding the properties of dark halos \citep{G72, P74, B91}.  
However, with the exception of \citet{Z70} and \citet{GSS89},
a framework for describing the properties and dynamics 
of filamentary and sheet-like structures in the cosmic 
web is still lacking.   
The main goal of this paper is to provide such a framework:  our 
goal is to discuss the cosmic web using language which extends 
naturally that currently used for discussing dark halos.  This 
complements recent work showing how the same framework can be used 
to discuss cosmic voids \citep{SvdW04}.  

In current descriptions of large scale structure, the abundance by 
mass of dark matter halos and the evolution of this abundance, 
i.e., the cosmological mass function $n(M,z)$, plays a fundamental 
role.  Models of this mass function suggest that it encodes information
about both gravitational dynamics and the statistics of the initial 
fluctuation field \citep{P74, B91}.  Models of $n(M,z)$ which 
assume that dark halos form from a spherical collapse are in 
reasonable but not perfect agreement with results from numerical 
simulations of hierarchical gravitational clustering \citep{S99}.  
In reality, halos are triaxial \citep{J02}, so they cannot have
formed from a spherical collapse.  The initial shear field and 
tidal effects almost certainly play some role in determining the 
evolution of an object, and models of the associated non-spherical 
collapse have been developed \citep{Z70, I73, W79, P80, B96}.  
Such non-spherical collapse models can be incorporated into models 
of the halo mass function in various ways \citep{EL95, M95, L98, C01}.  
In particular, \citet{SMT01} argued that much of the discrepancy 
between model predictions and simulations is removed if one replaces 
the assumption of a spherical collapse with one where halos form from 
a triaxial collapse.  

In this paper, we exploit a crucial difference between spherical 
and ellipsoidal collapse models of halo formation.  Namely, if we 
think of a halo as being triaxial, then the formation of an object 
corresponds to the time when gravity has caused all three axes to 
turn around from the universal expansion and collapse \citep{SMT01}.  
If the object is spherical, this collapse occurred at the {\em same} 
time for all three axes.  In contrast, a triaxial object has three 
critical times, corresponding to the collapse along each of the 
three axes---the shortest axis collapses first, the intermediate 
axis later and the longest axis last.  

If we identify `halos' with objects which have collapsed along all 
three axes, then it is natural to identify filaments with objects 
which have collapsed along only two axes, and sheets with objects 
which have only collapsed along one axis.  Given this identification, 
the remainder of this paper describes how to construct a model which 
naturally encapsulates the idea that halos form within filaments 
which themselves populate sheets.  In particular, we provide a 
framework for discussing the mass functions of sheets and filaments, 
and how these evolve, as well as the conditional mass functions 
of halos within filaments and sheets, and their evolution.  

Section~\ref{model} presents our model.  
It begins with a discussion of our triaxial collapse model 
(Section~\ref{ecmodel}).  
Appendix~A compares this model with others in the literature.  
It then translates this model into a form which is most convenient 
for use in the `excursion set' approach 
(Sections~\ref{excursionset}---\ref{hittingtimes}).  
The excursion set approach is commonly used to estimate the mass 
function of halos \citep{B91,SMT01}.  
Our collapse model, when combined with the excursion set approach, 
has a rich structure.  We illustrate this by using it to generate 
analytic approximations to the mass functions of sheets and filaments 
(Section~\ref{nmz}), and to the conditional distributions of 
filaments in sheets, and halos in filaments (Section~\ref{cmf}).  
Comments on how mass accretion may depend on environment 
(Section~\ref{accrete}), and on the characteristic densities of 
sheets, filaments and halos, which may aid identification of 
such objects in the cosmic web (Section~\ref{id}) conclude this 
section.  
Section~\ref{discuss} summarizes our findings.  

\section{A model for sheets, filaments and halos}\label{model}

\subsection{Ellipsoidal collapse}\label{ecmodel}
The gravitational collapse of homogeneous ellipsoids was studied 
by \citet{I73}, \citet{W79}, and \citet{P80}.  Although the exact 
evolution must be solved-for numerically, \citet{W79} provide an 
elegant analytic approximation for the evolution which is 
remarkably accurate.  
Unfortunately, these early analysis did not reduce, as they should, 
to the Zeldovich approximation in the linear regime.  
\citet{B96} noted that this was because they did not include 
the effects of the external tide self-consistently; once these 
effects are included, the collapse model is indeed self-consistent.  
Appendix~A provides details, and also shows how the analysis of 
\citet{W79} can be extended to this self-consistent case.  
In this model, the initial collapse is expected to be dominated 
by the local strain tensor, which includes both internal and 
external tidal forces.  For any given cosmology, the evolution of 
a patch in this model is determined by three quantities:  
its initial overdensity, $\delta$, and two shape parameters, 
$e$ and $p$.  It is this model which we will use in what follows.

\subsection{Mass functions and random walks:  
Three barriers for three axes}
\label{excursionset}
In the `excursion set' approach \citep{B91}, an approximation for 
the mass fraction in bound virialized halos of mass $m$ is obtained 
by mapping the problem to one which involves the first crossing 
distribution of a suitably chosen barrier by Brownian motion random 
walks.  The choice of barrier is set by the collapse model and by 
the epoch for which one wishes to estimate the mass function, and 
the mapping between random walk variables and halo masses is set by 
the shape of the power spectrum of the initial fluctuation field.  

In particular, if $f(\nu)$ denotes the fraction of walks which first 
cross a barrier at scale $\nu$, where $\nu$ is the random walk 
variable, then the mass fraction in objects of mass $m$ is 
\begin{equation}
 \nu f(\nu) \equiv 
  m^2 \frac{n(m,z)}{\rho_{m_0}}\frac{d\ln m}{d\ln \sigma^2}
\label{nufnu}
\end{equation} 
\citep{B91}.  Here $\sigma^2(m)$ is the variance in the initial density 
fluctuation field, smoothed on scale $R = (3m/4\pi\rho_{m_0})^{1/3}$, 
and extrapolated using linear theory to the present time.  

If the collapse is spherical, then the barrier is particularly simple:  
it has constant height, with higher barriers required to model the 
halo population at higher redshift.  In particular, this height is 
very simply related to the initial overdensity required for spherical 
collapse by redshift $z$.  This value, which is usually denoted 
$\delta_{sc}(z)$, will play a special role in what follows.  

In the ellipsoidal collapse model, the barrier height associated 
with halo formation by redshift $z$ depends on three numbers:  
the initial overdensity, and the shape parameters $e$ and $p$.  
In principle, then, the problem of estimating the mass function 
is one of crossing a barrier in a higher dimensional space 
\citep{SMT01, C01, ST02}.  \citet{SMT01} suggested that the 
computational complexity could be reduced significantly if one 
used suitably chosen representative values of $e$ and $p$.  
Once these have been set (for reasons given in Appendix~A, 
they suggested $p=0$ and $e = [\sigma(m)/\delta]/\sqrt{5}$), 
the excursion set approach corresponds to finding the first crossing 
distribution by a {\em one-dimensional} random walk in $\delta$ of a 
barrier whose height is {\em mass-dependent}.  
In this respect, the estimation of the mass function associated 
with ellipsoidal collapse proceeds exactly as for spherical 
collapse---the only complication is that the barrier height is no 
longer constant.  

Our model for the mass function in filaments and sheets is 
entirely analogous.  In principle, filaments at $z$ can form 
from some combination of $\delta$, $e$ and $p$; the precise 
combination depends on the details of the collapse model.  
In our preferred model, the same one used by \citet{SMT01}, 
this requires that collapse occurs along the intermediate 
length axis at $z$, but along the longest axis at a later time.  
Thus, estimation of the mass fraction in filaments requires solution 
of a high-dimensional random walk, and, as for collapse along all 
three axes, we approximate by setting $e$ and $p$ equal to 
representative values, and considering a one-dimensional walk in 
the remaining variable $\delta$.  
In this case, the barrier height associated with 
filament formation is that associated with collapse along two, 
rather than three axes.  And a similar argument means that the 
mass function in sheets can be approximated by studying the first 
crossing distribution of the one-dimensional barrier associated 
with collapse along just one axis.  

\begin{figure}[!h]
\begin{center}
\plotone{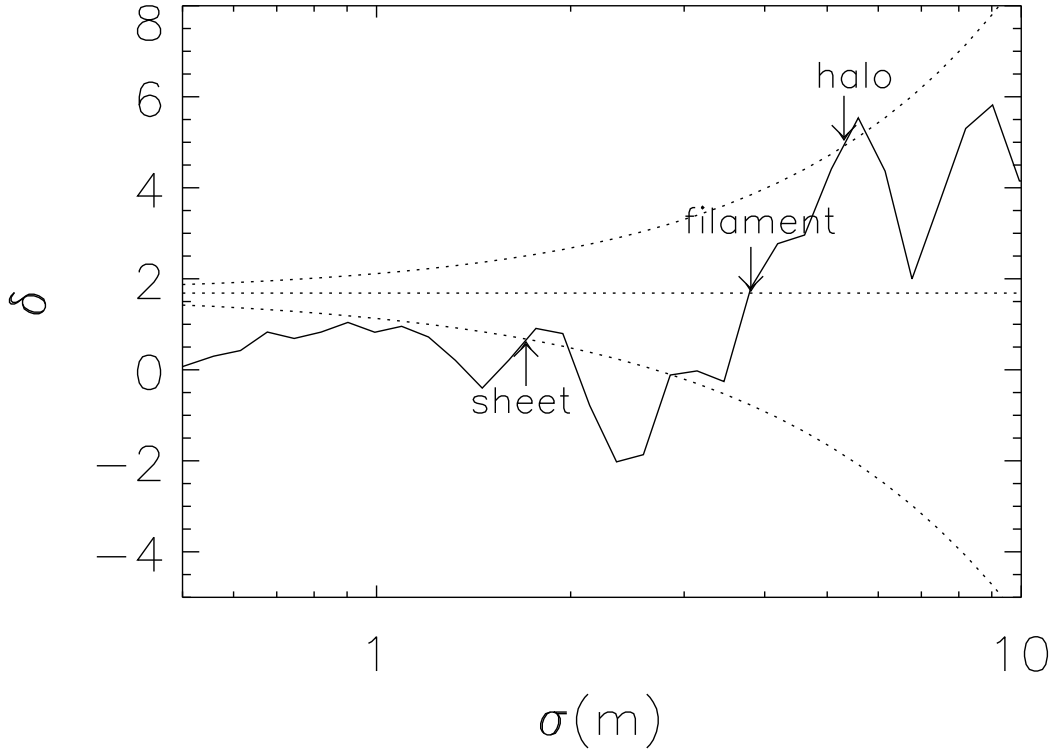}
\end{center}
\caption{An example of a random walk (solid line) crossing the 
   barriers (dotted lines) associated with sheets, filaments
   and halos (bottom to top). 
   The fraction of walks which first cross the lowest barrier 
   at $\sigma(m_s)$, then first cross the second barrier at 
   $\sigma(m_f)$ and finally cross the highest barrier at 
   $\sigma(m_h)$ represents the mass fraction in halos of 
   mass $m_h$ which are embedded in filaments of mass $m_f>m_h$, 
   which themselves populate sheets of mass $m_s>m_f$ (recall 
   that $\sigma$ is a decreasing function of $m$).  
   The precise barrier shapes depend on the collapse model; 
   the dotted curves show the barriers in equation~(\ref{movingbarriers}).}
\label{randomwalk}
\end{figure}

Figure~\ref{randomwalk} illustrates the basic idea.  
The jagged line shows a 
random walk, and the three dotted lines show the barriers associated 
with collapse along one (bottom), two (middle) and three (top) barriers.  
In our model, the fraction of walks which first cross the lowest 
barrier on scale $\sigma(m)$ equals the mass fraction in sheets of 
mass $m$, the fraction which first cross the second barrier at 
$\sigma(m)$ equals the mass fraction in filaments of mass $m$, 
and the fraction which first cross the highest barrier at $\sigma(m)$ 
equals the mass fraction in triaxial halos of mass $m$.  We will 
provide estimates of these quantities shortly.  

However, Figure~\ref{randomwalk} shows that this model can also be used 
to provide significantly more information.  For instance, notice that 
all walks first-cross the barrier associated with sheets at a higher 
mass scale than when they first-cross the barriers associated with 
filaments or halos.  Thus, in addition to providing a way to estimate 
the mass functions of sheets, filaments and halos, our approach also 
provides a framework for discussing the mass fraction in halos of 
mass $m_h$ that are embedded in filaments of mass $m_f<m_s$ which are 
themselves surrounded by sheets of mass $m_s>m_f$.  
Figure~\ref{randomwalk} shows that this is simply related to the 
fraction of walks which are conditioned to pass through certain points 
before first crossing $m_h$.  Thus, our approach allows one to estimate 
how halo abundances correlate with the {\em morphology} and density 
of the large-scale structure which surrounds them.  

So far, we have only discussed the model at a fixed redshift.  
Since the barrier shapes depend on redshift, our approach also allows 
one to estimate how sheet, filament and halo abundances evolve.  
Moreover, it allows one to address questions like:  is the halo 
abundance in sheets of mass $10^{15} M_\odot$ at $z=1$ any 
different from the mix of halos in $10^{15} M_\odot$ filaments 
today?  Our model naturally incorporates the fact that the mass in 
high redshift sheets becomes partitioned up among filaments at later 
times, with filaments themselves being partitioned into halos.  
Clearly, the model is rich in spatial, temporal and spatio-temporal 
information.

\subsection{Mass dependence of the three barriers}\label{3barriers}

The key output from the triaxial collapse models is an estimate 
of the typical overdensity required for collapse along one, two 
and three axes by redshift $z$.  
The dotted curves in Figure~\ref{randomwalk} show how these three 
`barriers' depend on mass.  From bottom to top, the curves show 
\begin{eqnarray}
 \delta_{ec1}(\sigma , z) & = & \delta _{sc} (z)
  \left \{ 1-0.56\left[{{\sigma ^2} \over {\delta _{sc}^2 (z)}} \right] ^{0.55}
   \right \},  \nonumber \\
 \delta_{ec2}(\sigma , z) & = & \delta _{sc} (z)
  \left \{ 1-0.012\left[{{\sigma ^2} \over {\delta _{sc}^2 (z)}}\right] ^{0.28}
   \right \} \approx \delta _{sc} (z),
   \label{movingbarriers} \\
 \delta_{ec3}(\sigma , z) & = & \delta _{sc} (z)
  \left \{ 1+ 0.45 \left[{{\sigma ^2} \over {\delta _{sc}^2 (z)}}\right]^{0.61}
   \right \}. \nonumber
\end{eqnarray}
These analytic approximations to the barriers associated with 
collapse along one, two and three axes are reasonably accurate 
(c.f. Figure~\ref{shethall}).  
They show that the critical overdensity for ellipsoidal 
collapse along all three axes, $\delta_{ec3}$, is larger than in 
the spherical collapse model (for which this number is $\delta_{sc}$).  
However, collapse along just two axes requires an overdensity 
$\delta_{ec2}$ which is almost exactly that in the spherical model, 
and collapse along one axis only requires a smaller initial 
overdensity, $\delta_{ec1}$.  
In this model, tidal forces enhance collapse along the first 
axis and delay collapse along the last axis relative to the spherical 
collapse model \citep{SMT01}---the expressions above quantify these 
effects. Note that the differences among the three critical 
overdensities are larger for larger values of $\sigma$, 
corresponding to ellipsoids of lower masses.    

Notice that these barrier shapes depend both on $\sigma(m)$  
and on $\delta_{sc}(z)$. 
The presence of these two terms reflects the fact that 
the collapse depends on the expansion history of the universe, 
and on the initial spectrum of fluctuations.  What is of particular 
importance in what follows is that the barrier shapes actually 
depend, not on $\delta_{sc}$ and $\sigma$ individually, but on 
the scaling variable  $\nu=[\delta_{sc}(z)/\sigma(m)]^2$.  
In the excursion set approach, this implies that the mass functions 
of sheets, filaments and halos at any given time, in any given 
cosmology, and for any given initial fluctuation spectrum, can all 
be scaled to universal functional forms.  Our next step is to provide 
analytic approximations to these forms.  

\subsection{First crossing distributions}\label{hittingtimes}
\begin{figure}[!ht]
\begin{center}
\plotone{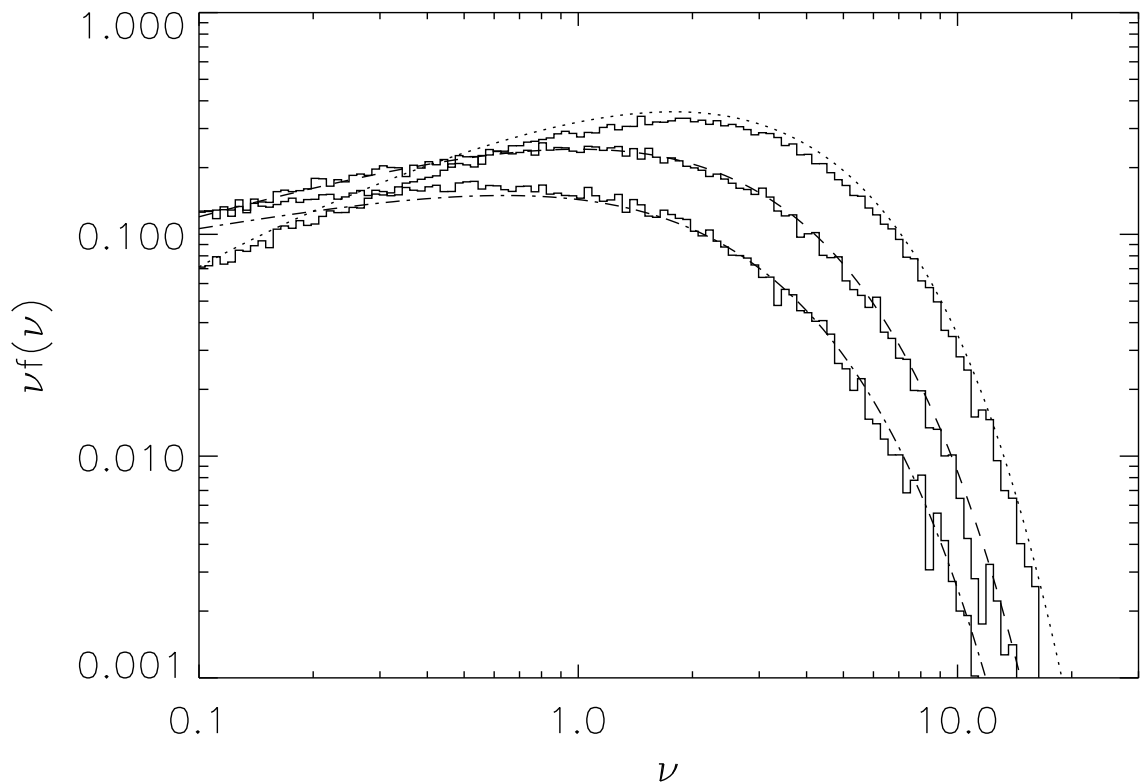}
\end{center}
\caption{Histograms show the first crossing distributions of the 
         barriers associated with collapse along one, two and 
         three axes (histograms which extend to largest $\nu$ are 
         those for collapse along only one axis).  Curves show 
         analytic approximations to these distributions following 
         \citet{ST02}.  }
\label{fcross}
\end{figure}

In the simplest excursion set model, object abundances are related 
to the first crossing distributions by one-dimensional random walks 
of the moving barriers in equation~(2).  These 
distributions are easily obtained by Monte-Carlo simulation.  
(Alternative numerical methods are also available, e.g., Zhang \& Hui 2006.)  
The Monte-Carlo results are simplified considerably by the fact that 
the barriers can be expressed in terms of the scaling variable 
 $\nu=[\delta_{sc}(z)/\sigma(m)]^2$.  
The histograms in Figure~\ref{fcross} show the results.  
The abundances of sheets, filaments and halos are related to these 
first crossing distributions by equation~(\ref{nufnu}).  

The smooth curves show analytic approximations to these first 
crossing distributions, computed following \citet{ST02}.  
Namely, when the barrier is of the form 
\begin{equation}
 \delta_{ec}(\sigma, z) = \delta _{sc} (z) \,
                     \Bigl[1+ \beta \nu ^{-\alpha}\Bigr],  \nonumber
 \label{barrier}
\end{equation}
as is the case for all three barriers in equation~(\ref{movingbarriers}), 
then the first crossing distribution is approximately 

\begin{equation}
 \nu f(\nu) = \sqrt{\frac{\nu}{2\pi}}e^{-\nu[1+\beta \nu^{-\alpha}]^2/2} 
 \left\{1+\frac{\beta}{{\nu}^{\alpha}} \left[ 1- \alpha +
        \frac{\alpha(\alpha -1)}{2!} + \dots \right] \right\} .
 \label{approx}
\end{equation}
Alternative analytic estimates could be computed by noting that 
when $\alpha=1/2$, then the first crossing distribution can be 
written analytically in terms of sums of parabolic cylinder functions 
\citep{B67}.  Given that these would only provide approximations 
to the first crossing distributions we require (because $\alpha\ne 1/2$), 
we have chosen the considerably simpler approximation given in 
equation~(\ref{approx}).  Figure~\ref{fcross} indicates that these 
approximate solutions are sufficiently accurate to allow analytic 
estimates of a number of interesting quantities.  

\begin{figure}[!ht]
\begin{center}
\plotone{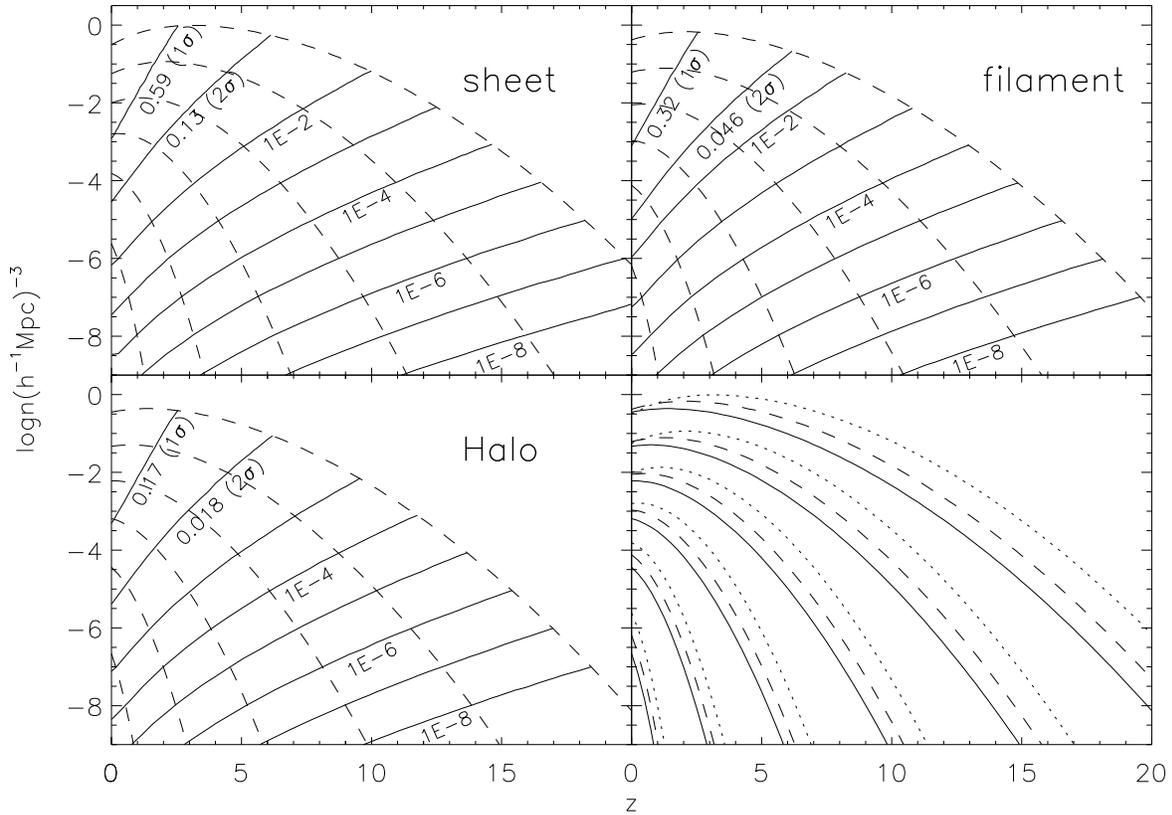}
\end{center}
\caption{The upper left, upper right and bottom left are the comoving 
         number densities (dashed lines) and mass fractions (solid lines) 
         for sheets filaments and halos respectively. The bottom right 
         of the figure compares the comoving number densities of sheets 
         (dotted lines), filaments (dashed lines) and halos (solid lines). 
         The top sets of curves are for objects with masses of 
         $10^{10} M_{\odot}$, and subsequent curves show masses 
         which are larger by 1~dex until the bottom sets of curves, 
         which are for masses of $10^{15} M_{\odot}$.}
\label{numfracall}
\end{figure}

\subsection{Abundance by mass and redshift}\label{nmz}
Upon assumption of a cosmology [which determines $\delta_{sc}(z)$] 
and a fluctuation spectrum [which determines $\sigma(m)$], 
the first crossing distributions shown in Figure~\ref{fcross} 
can be converted to mass functions using equation~(\ref{nufnu}). 
In what follows, we assume a spatially flat model with cosmological 
constant $\Lambda_0 =1-\Omega_0$, where 
 $(\Omega_0,h,\sigma_8) = (0.3,0.7,0.9)$.  

The dashed lines in the bottom-left panel of Figure~\ref{numfracall} 
show the evolution of the number densities of halos having mass 
$10^{10} M_{\odot}$ (top) to $10^{15} M_{\odot}$ (bottom).  
Except for the lowest mass range, these curves decrease monotonically 
with increasing redshift. This reflects the well-known hierarchical 
nature of structure formation:  at any given redshift, massive halos 
are less common than halos of lower mass, and they are even less 
common at earlier times.  
As a result, if one integrates the mass fraction in objects above 
some minimum mass, then this fraction is smaller at higher redshift.  
If, instead, one fixes the mass fraction, then one must integrate down 
to smaller mass halos to obtain this fraction.  Since less massive 
halos are more abundant, plots of the abundance associated with a 
fixed mass fraction increase with increasing redshift 
\citep{MW02}.   
The solid curves show such locii in the abundance-redshift plane 
for a range of mass fractions.  
The top-left and right panels show analogous results, but for sheets 
and filaments respectively.  Similar trends are seen in all three 
panels, indicating that sheets and filaments grow hierarchically 
in much the same way as halos.  

These panels show that, at $z=0$, more than half of the cosmic mass 
is in sheets with masses exceeding $10^{13}M_{\odot}$.
This fraction is about $30\%$ for filaments in the same mass range, 
and is only $13\%$ for halos.  
At the present time, more than $99\%$ of the cosmic mass
is contained in sheets more massive than $10^{10}M_{\odot}$, in 
contrast to virialized halos which, in this mass range, contain only 
about half of the cosmic mass.  
At $z\sim 3$, more than 10\% of the cosmic mass was already 
assembled in sheets more massive than that of the Milky Way halo 
($\sim 10^{12}M_\odot$), while only $2\%$ was assembled to halos in 
this mass range. Even at $z\sim 10$, more than $1\%$ of the cosmic 
mass was already in sheets more massive than $10^{10}M_\odot$, 
whereas the mass fraction in halos in the same mass range was ten 
times smaller.  This suggests that the structure at high $z$ is 
dominated by sheets.  

At any given redshift, we may define a characteristic mass, $M_*(z)$, 
so that $\sigma(M_*)$ is equal to the critical overdensity 
$\delta_{sc}(z)$.  In our $\Lambda$CDM model, the value of $M_*$ drops rapidly 
with redshift, from just over $10^{13}M_{\odot}$ at $z=0$ to 
$\sim 10^{10.5}M_{\odot}$ at $z=2$.  Objects at $z$ whose mass 
satisfies $\sigma(M)=N\delta_{sc}(z)=N\sigma(M_*)$, where 
$N=1,2,\cdot\cdot\cdot$, can be thought of as being increasingly 
unusual compared to objects of the characteristic mass $M_*(z)$.  
The top two solid curves in the first three panels of 
Figure~\ref{numfracall} show the abundance-redshift locii of 
sheets, filaments and halos whose masses satisfy 
$\sigma(M)=\sigma(M_*)$ and $\sigma(M)=2\sigma(M_*)$.  
The panels show that the mass fraction in $1\sigma$ sheets, $0.59$, 
is larger than that in $1\sigma$ filaments (0.32) and than that 
in $1\sigma$ halos (0.17).  A similar scaling applies to the more 
massive $2\sigma$ objects, implying that, at any redshift, 
the typical structure is dominated by sheets.

The bottom right panel in Figure~\ref{numfracall} compares the number 
densities of sheets (dotted), filaments (dashed) and halos (solid) at 
different redshifts in the three other panels directly (i.e., these 
were the dashed curves in the other three panels):  
top-most curves show our model for masses of $10^{10} M_\odot$, 
bottom-most for $10^{15} M_\odot$, and the curves in between 
show results in which the mass changes by 1~dex.  
These curves show that, at masses larger than $10^{12} M_{\odot}$, 
there are more sheets than filaments, and more filaments than halos, 
for almost all redshifts.  However, at small masses and late times 
this trend begins to reverse:  at $z=0$ there are fewer sheets than 
filaments with masses smaller than $10^{11}  M_{\odot}$, 
and there are more halos than sheets of mass $10^{10}  M_{\odot}$.  
Based on these mass functions, we can estimate the average mass 
of objects more massive than some given minimum mass.  For instance, 
when the minimum mass is $10^{10}  M_{\odot}$ then, at the 
present time, this average mass is
$3.9 \times 10^{11}  M_{\odot}$ for sheets, 
$2.1 \times 10^{11}  M_{\odot}$ for filaments, and  
$1.6 \times 10^{11}  M_{\odot}$ for halos.
The average mass of sheets is about 2.5 times that of halos, 
again suggesting that large scale structure is dominated by 
sheets rather than by virialized halos.

\begin{figure}[!ht]
\begin{center}
\plotone{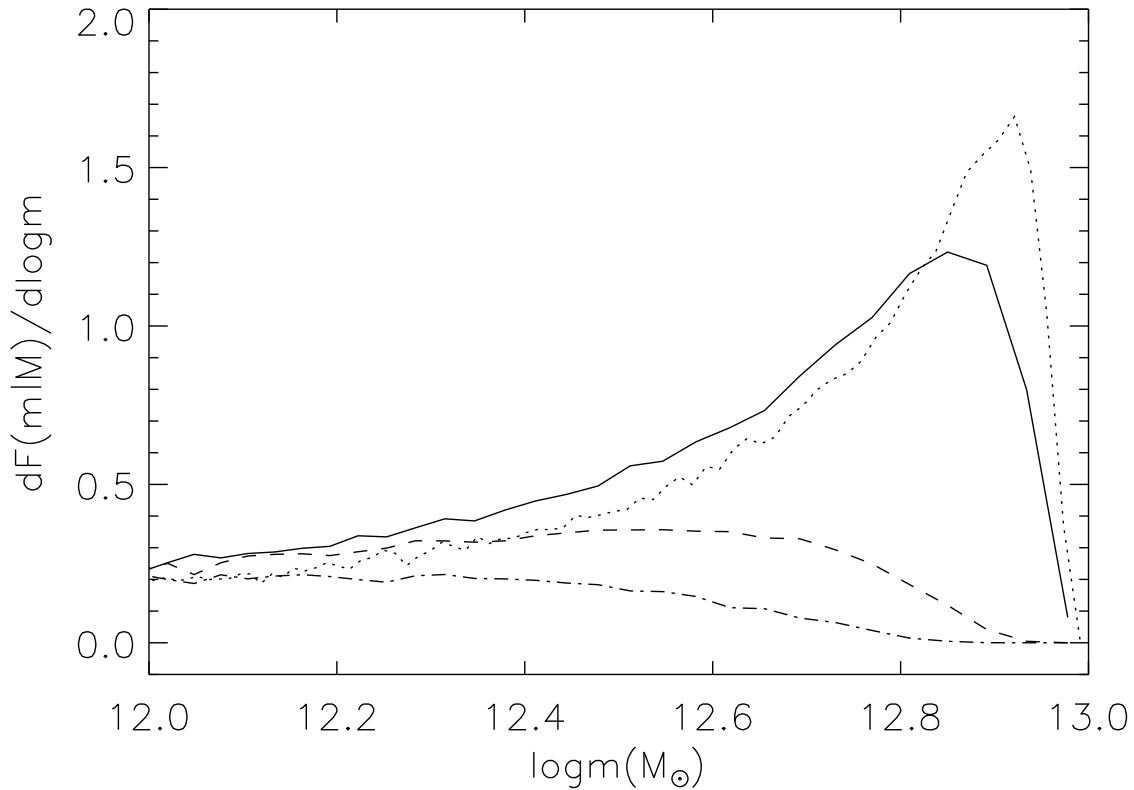}
\end{center}
\caption{Mass fraction of $10^{13} M_{\odot}$ sheets that is 
         in filaments (solid) and halos (dashed) of mass $m$ all 
         at $z=0$.  Dotted curve shows the mass fraction of 
         $10^{13}M_{\odot}$ filaments at $z=0$ that is in halos.
	 Dot-dashed curve shows the mass fraction contained in halos 
	 of mass m within an average volume of the universe of 
	 the same mass  ($10^{13} M_{\odot}$).   
         The differences between the dotted and dashed curves indicate 
         that, at fixed large-scale overdensity, the halo population 
         is expected to be correlated with the morphology of the 
         surrounding large-scale structure. }
\label{cosmicweb}
\end{figure}

\subsection{Conditional mass functions}\label{cmf}
Halo abundances are expected to correlate with the overdensity 
of their surroundings---massive halos populate dense regions 
\citep{MW96}.  However, our model predicts that halo abundances 
will also correlate with the morphology of their surroundings.  
To illustrate, the solid curve in Figure~\ref{cosmicweb} shows the 
mass fraction of $10^{13}  M_{\odot}$ sheets that is in 
filaments of mass $m$ at $z=0$.  
The figure indicates that the most probable filament mass, 
$7\times 10^{12} M_{\odot}$, is a substantial fraction of that 
of its parent sheet.  Although we do not show it, the analytic 
estimate of this quantity, given by inserting our barriers for 
filaments and sheets in equation~(7) of \citet{ST02}, is in 
excellent agreement with this distribution.  
This solid curve should be compared with the dashed one, which shows 
the mass fraction in halos (rather than filaments) of mass $m$ that 
are known to be in sheets of mass $10^{13}  M_{\odot}$ at $z=0$.  
(In this case, the analytic estimate is not as accurate.)
Notice that the most probable halo mass is significantly smaller 
than that of the parent sheet, and the halo mass function in sheets 
is skewed significantly towards lower masses than that of filaments.  
As a final comparison, the dotted line shows the mass fraction 
of $10^{13}  M_{\odot}$ filaments (rather than sheets) that is 
in halos of mass $m$, and the dot-dashed curve shows the mass fraction 
contained in halos of mass m within an average volume of the universe of 
the same mass  ($10^{13} M_{\odot}$). 
Here again, the mass function within sheets and filaments is biased
compared to the average universe and 
the most probable halo mass is a substantial fraction 
of the mass of the parent filament.  The significant differences 
between the dotted and dashed curves illustrate an important 
physical implication of our model:  if the properties of a galaxy 
are correlated with the mass of its host halo, then the galaxy 
population in sheets of a given mass is expected to be different 
from that in filaments of the same mass.  In other words, at fixed 
large-scale mass, galaxy properties are expected to be correlated 
with the morphology of the surrounding large scale structure.

\begin{figure}[!ht]
\begin{center}
\plotone{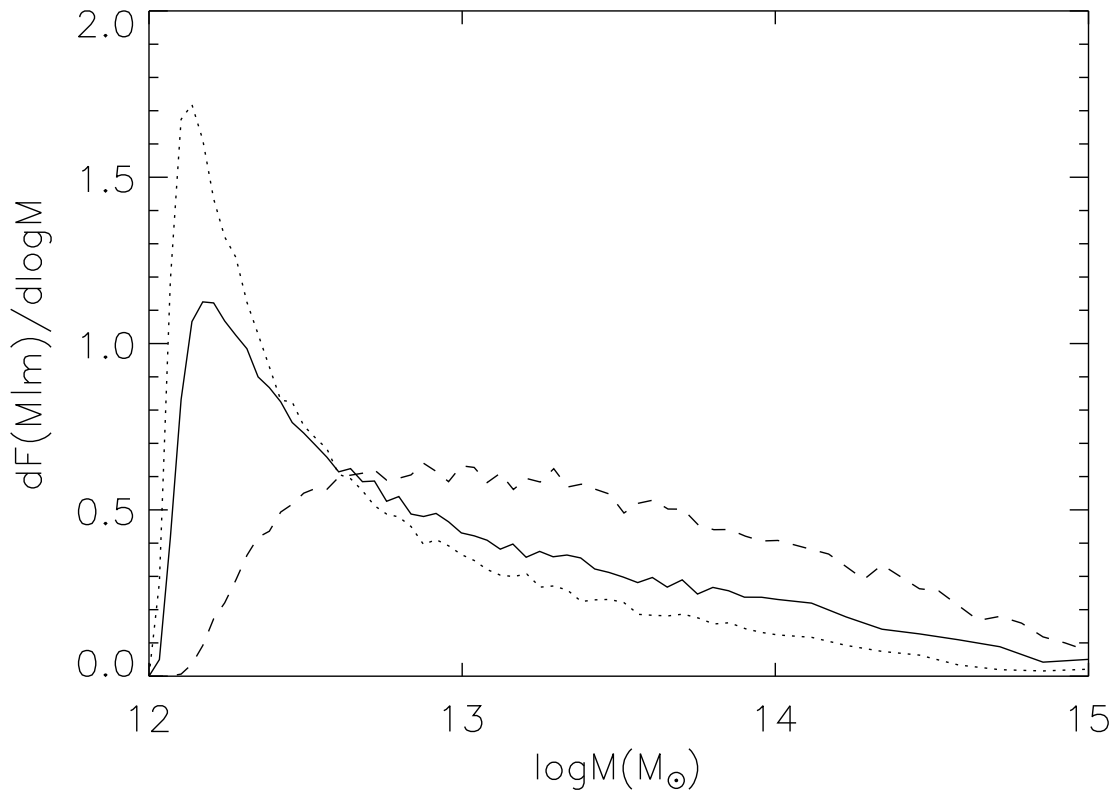}
\end{center}
\caption{Conditional distribution of sheets centred on a smaller 
         scale filament of mass $10^{12} M_{\odot}$ (solid).  
         Dotted and dashed curves show the distribution of 
         filaments and sheets which surround $10^{12} M_{\odot}$ 
         halos.}
\label{cosmicweb2}
\end{figure}

Figure~\ref{cosmicweb} shows the distribution of masses of smaller 
scale objects when the large scale mass and morphology is fixed.  
But our model also allows one to estimate the mass and morphology 
of the environment which surrounds a given smaller scale structure.  
To illustrate, Figure~\ref{cosmicweb2} shows conditional mass 
functions in which the small-scale object is fixed, and we study 
the distribution of the surrounding large scale mass and morphology.  
The solid curve shows the distribution of sheets around 
$10^{12} M_{\odot}$ filaments, and the dotted and dashed 
curves show the distributions of filaments and sheets around 
$10^{12} M_{\odot}$ halos.  (Once again, the analytic estimates 
of these distributions are in reasonable but not perfect agreement.)  
Notice again that the mass of a sheet is well correlated with the 
mass of the filament it surrounds, as are the masses of filaments 
with those of the halos they surround.  On the other hand, halo and 
sheet masses can be very different.  
These scalings indicate that our model can be used to estimate how 
the surrounding large scale structure correlates with different 
galaxy populations.

\subsection{Filamentary versus isotropic mass growth}\label{accrete}
Our model can be used to provide a simple estimate of how halos 
grow.  To see why, consider a random walk with height equal to 
the rms fluctuation:  $\sigma$.  
The typical mass scales of sheets, filaments and halos are given 
by setting this value equal to the various barrier heights: 
 $\sigma = \delta_{\rm ec1,2,3}(\sigma,z)$.  
To see what this implies, approximate the three moving barriers 
in equation~(2) as 
\begin{eqnarray*}
  \delta_{\rm ec1}(\sigma,z) &\approx& \delta_{\rm c}(z)
     \left\{1 - 0.5 {\sigma\over \delta_{\rm c}(z)}\right\} 
     = \delta_{\rm c}(z) - 0.5\,\sigma;\nonumber\\
  \delta_{\rm ec2}(\sigma,z) &\approx& \delta_{\rm c}(z); \nonumber\\
  \delta_{\rm ec3}(\sigma,z) &\approx& \delta_{\rm c}(z)
     \left\{1 + 0.5 {\sigma\over \delta_{\rm c}(z)}\right\} 
     = \delta_{\rm c}(z) + 0.5\,\sigma .
\end{eqnarray*}  
Then the typical mass scale of filaments is 
 $\sigma_{\rm f} \approx \delta_{\rm c}(z)$.  
Similarly, a characteristic mass scale for sheets is 
 $\sigma_{\rm s} \approx \delta_{\rm c}(z) - 0.5\, \sigma_{\rm s}$,
implying 
 $\sigma_{\rm s} = (2/3)\,\sigma_{\rm f}$, 
whereas the mass scale of halos is 
 $\sigma_{\rm h} = 2\,\sigma_{\rm f}$.  

Massive objects are associated with more extreme fluctuations.  
We can approximate by studying the case in which the walk height 
is twice the rms value (a $2\sigma$ fluctuation).  
In this case, $\sigma_{\rm f} = \delta_{\rm c}(z)/2$, 
$\sigma_{\rm s} = (2/5)\,\delta_{\rm c}(z) = (4/5)\,\sigma_{\rm f}$, 
and  $\sigma_{\rm h} = (2/3)\,\delta_{\rm c}(z) = (4/3)\,\sigma_{\rm f}$.  
Notice that, for massive objects, these factors are substantially 
closer to unity than they are for more typical objects.  
This suggests that the most massive halos are a substantial fraction 
of the mass of the filaments they populate, and these filaments are 
a substantial fraction of the sheets in which they are embedded.  
Thus, massive halos will appear to be accreting most of their mass 
from filaments.  In contrast, a lower mass halo may be substantially 
less massive than the filament it inhabits.  The growth of such a 
halo will appear to be dominated by mergers with the other small 
halos which populate the same filament (and possibly the other nearby 
filaments).  
In this respect, our model is qualitatively consistent with 
observations which suggest that massive clusters grow by accretion 
along filaments, whereas the growth of lower mass objects is less 
obviously anisotropic.   

\subsection{Identification of sheets, filaments and halos in the cosmic web}
\label{id}
Although our approach provides a framework for discussing the 
morphology of the cosmic web, we have, so far, not made any 
statements about precisely how the sheets, filaments and halos 
in our formalism are to be identified in cosmic density fields.  
Our collapse model is calibrated so that the object which forms 
after collapse has been completed along all three axes, i.e., a 
halo, has the same density relative to the background as is 
expected in the spherical collapse model.  Sheets in our model 
should be approximately this overdensity to the one-thirds power, 
whereas filaments are this to the two thirds power.  
Although, formally, the density of a halo depends on the 
background cosmological model, it is common practice to identify halos 
at a given epoch as objects which are 200 times denser than the 
critical density at that epoch.  Therefore, filaments and sheets 
should be approximately 36 and 6 times denser than the critical 
density.  This provides a simple rule of thumb for identifying 
sheets and filaments in the dark matter density field.  
Note that, in the context of our model, it may be more
appropriate to think of massive halos as lying at the centers
of filaments, rather than defining their endpoints.
It will be interesting to see how these simple estimates
compare with measurements in simulations.

\section{Discussion and Conclusion}\label{discuss}
We extend the excursion set approach to quantify how the cosmic web 
is made up of sheets, filaments and halos.  
Our model assumes that objects form from a triaxial collapse; 
we define sheets as objects which have collapsed along only one axis, 
filaments as objects which have collapsed along two axes, and halos 
as objects where all three axes have collapsed.  Therefore, our 
model requires specification of exactly how triaxial collapse occurs.  
Appendix~A discusses our preferred collapse model, compares it with 
Zeldovich's approximation, and shows how the analytic arguments of 
\citet{W79} can be extended to provide an accurate analytic description 
of our ellipsoidal collapse model.  

The details of how a patch collapses depends on its initial overdensity 
$\delta$, and on its initial shape parameters $e$ and $p$.  To embed 
this collapse model in the excursion set approach requires study of 
three-dimensional random walks crossing a barrier $B(\delta,e,p)$.  
Here, we follow \citet{SMT01} and study the simpler problem in which 
$e$ and $p$ are replaced by representative values, and then study a 
one-dimensional boundary crossing problem.  We emphasize that this 
is only an approximation, albeit a useful one.  

For any redshift $z$, insertion of the representative values of $e$ 
and $p$ in our collapse model provides estimates of the critical 
overdensities required for collapse along one, two and three axes.  
We find that these overdensities, 
$\delta_{ec1}(z,\sigma)$, $\delta_{ec2}(z,\sigma)$ and 
$\delta_{ec3}(z,\sigma)$ depend on both time and mass 
(equation~ 2).  Because of the dependence on 
$\sigma$, in the language of the excursion set approach, each of 
these critical overdensities is a `moving barrier' 
(Figure~\ref{randomwalk}).  Insertion of each moving barrier into 
the excursion set approach provides estimates of the mass fraction in 
sheets, filaments and halos as a function of mass and time 
(Figure~\ref{fcross}).  

In our model, halos of a given mass $m_h$ populate filaments which 
are more massive, and the filaments themselves are surrounded by even 
more massive sheets.  Hence, the characteristic masses of sheets are 
predicted to be substantially larger than of filaments or halos.  
Every halo at a given time was previously a filament of the same mass, 
and before that, a sheet.  A halo of mass $10^{13} M_\odot$ 
today was a filament at redshift $z\approx 0.45$ and a sheet at 
$z\approx 1$ (c.f. equation  2).  
Halo abundances are expected to correlate with the overdensity of 
their surroundings (massive halos populate dense regions).  
Our model predicts that, at fixed large-scale overdensity, halo 
abundances will also correlate with the morphology of their 
surroundings (Figures~\ref{cosmicweb} and~\ref{cosmicweb2}).  
Therefore, in models where the properties of a galaxy are correlated 
with the mass and formation history of its host halo, our model 
provides a framework for describing correlations between galaxy 
properties and the morphology of large scale structure.  
For instance, distributions like those in Figure~\ref{cosmicweb} 
may be used to study if morphological structures in the galaxy 
distribution, like the SDSS Great Wall at $z\sim 0.08$, are unusual.  
And distributions like those in Figure~\ref{cosmicweb2} provide a 
framework for understanding if galaxies at high redshift 
form preferentially in sheets or filaments \citep{Mo_etal05}.  
Because our model exhibits various physically appealing features, 
we anticipate that it will provide a useful framework for quantifying 
the relation between galaxies, halos, and the cosmic web.  

\acknowledgments
We would like to thank Gerhard B\"orner for organizing a meeting 
at the Ringberg Castle, Tegernsee, and the staff of the Castle for 
their hospitality when this project began.  We thank Joerg Colberg 
and an anonymous referee for insightful comments and questions.
JS thanks the University of Massachusetts at Amherst for hospitality 
during the summer of 2004.  
This work was supported by the NSF under CAREER award AST-0239709 
to TA, and grant AST-0520647 to RKS.

\appendix
\section{Appendix}
This Appendix describes the triaxial collapse model we use in the 
main text.  Although the evolution must be solved numerically in 
general, we also discuss a reasonably accurate analytic approximation 
to the evolution, and show that it is considerably more accurate 
than the Zeldovich approximation.  

Let $A_k$ denote the scale factors for the three principal axes 
of the ellipsoid.  Then 
\begin{equation}
{{{d^2} A_k\over {dt^2}}} = -{4\pi G {\bar{\rho}} A_k 
  \left[{1 + \delta\over 3} +{{b^{'}_k} \over 2}\,\delta
  + {\lambda ^{'}_{{\rm ext} k}}\right]}
\label{ellipsoid}
\end{equation}
\citep{B96}, where $\bar\rho\propto a^{-3}$ is the mean density 
of the universe, 
\begin{equation}
 \delta  \equiv \frac{\rho - {\bar\rho}}{\bar\rho}
            ={{a ^3} \over {A_1A_2A_3}} - 1
\end{equation}
is the relative overdensity, and ${b^{'}_k}\delta /2$ and 
$\lambda ^{'}_{{\rm ext}k}$ denote the interior and exterior tidal forces.  
In particular, 
\begin{equation}
  b^{'}_k(t)=b_k(t) - {2 \over 3}, \quad {\rm where} \ 
  b_k(t) = A_1(t)A_2(t)A_3(t)\int_0^\infty {{d\tau} \over {[A^2_k(t) +\tau] 
                                     \prod ^3 _{m=1} [A_m^2(t)+\tau]^{1/2}}},
\end{equation}
and the linear approximation for the external tides is 
\begin{equation}
 {\lambda ^{'}_{{\rm ext}k}} (t) =\frac{a}{a_i} {\lambda ^{'}_{k}}(t_i)
  =\frac{a}{a_i} \left[{\lambda _{k}}(t_i)-{\delta_i\over 3}\right]\,,
\end{equation}
where $a_i=a(t_i)$.
Note that the $\lambda_k$s are the initial eigenvalues of the 
strain tensor; they are related to the initial density contrast 
$\delta_i$ and the shear ellipticity $e$ and prolateness $p$ by:
\begin{equation}
 \lambda _1 = (\delta _i /3)(1+3e+p), \ \ \ \
 \lambda _2 = (\delta _i /3)(1-2p), \ \ \ {\rm and} \ \ \ 
 \lambda _3 = (\delta _i /3)(1-3e+p).
\label{lamdaj}
\end{equation}

The initial conditions are set by Zeldovich approximation:
\begin{equation}
 A_k(t_i) = {a}_i[1-\lambda_{k}(t_i)], \ \ {\rm and} \ \ 
 \dot{A}_k(t_i) = H(t_i)A_k(t_i) -a_i H(t_i)\lambda_{k}(t_i).
\end{equation}
These expressions reduce to those of the spherical collapse model
if the two tidal force terms are not included.  Note that, 
as a result of the tidal forces, an initially spherical region 
will be distorted into a collapsing homogeneous ellipsoid.

The tidal gravitational forces accentuate the asymmetry of the 
ellipsoid, so that the collapse of the three axes can happen at 
very different times.  
The shortest axis will collapse first, followed by the intermediate 
axis and then by the longest axis.  Absent angular momentum, the 
collapse of any given axis continues to arbitrarily small sizes, 
a well-known feature of the spherical collapse model as well.  

\begin{figure}[!ht]
\begin{center}
\plotone{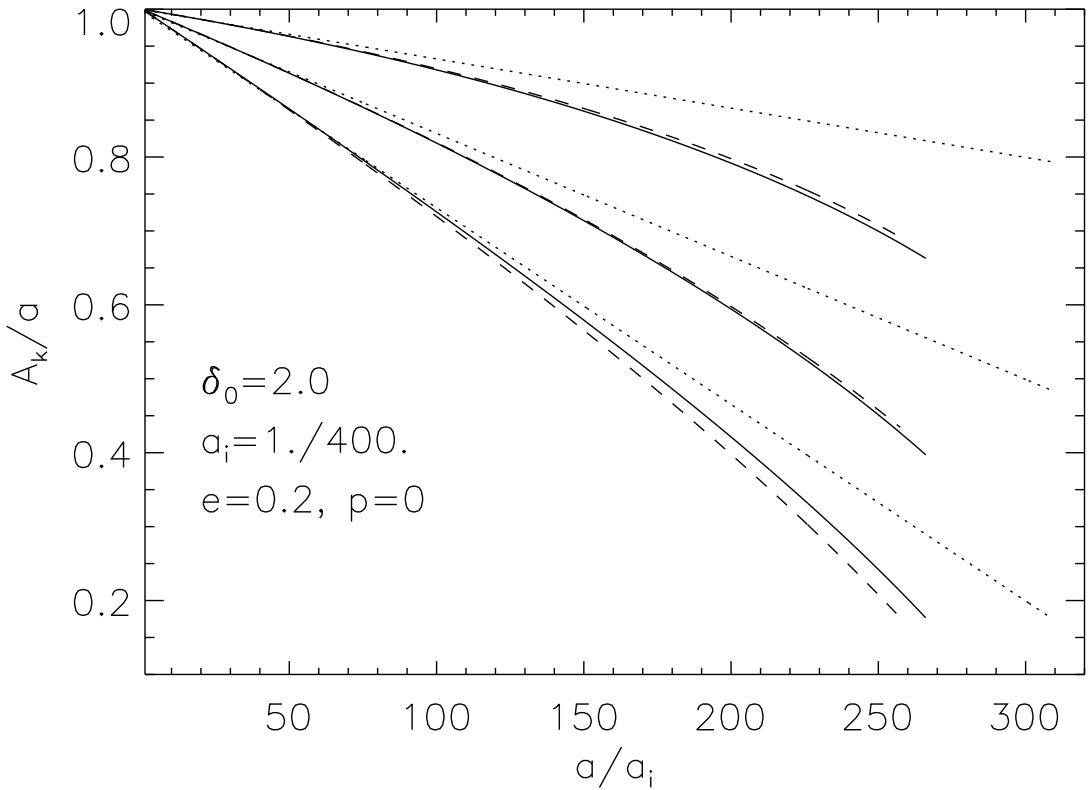}
\end{center}
\caption{Comparison of full ellipsoidal collapse model (solid), 
         Zeldovich approximation (dotted) and the approximation 
         in equation~(\ref{mowhitesilk}) for $e=0.2$ and $p=0$ 
         in an Einstein de-Sitter model. 
         The three sets of curves show the comoving lengths of 
         the shortest, intermediate and longest axes.  These 
         correspond to the formation of sheets, filaments and 
         halos.}
\label{bondmo}
\end{figure}

Given initial values of $\delta _i$, ellipticity $e$, prolateness $p$,
and epoch $a_i$, equation~(\ref{ellipsoid}) must be solved numerically 
for each axis $A_k$.  However, it is straightforward to extend the 
analytic approximation for ellipsoidal collapse provided by 
\citet{W79} so that it reduces self-consistently to the Zeldovich 
approximation at early times.  In particular, we write 
\begin{equation}
 A_k(t) \approx {a(t)\over a_i} A_k(t_i)\,\Bigl[1-D(t)\,\lambda_k\Bigr]
           - {a(t)\over a_i}A_h(t_i) \left[ 1-\frac{D(t) \delta_i}{3} 
                             - \frac{a_e(t)}{{a}(t)} \right],
 \label{mowhitesilk}
\end{equation}
where $A_h(t_i)=3/\sum_k A_k(t_i)^{-1}$, 
$D(t)$ is the linear theory growth mode, 
and $a_e(t)$ is the expansion factor of a universe with initial 
density contrast $\delta_i\equiv\sum_k \lambda_k(t_i)$.  
Note that the first term in the expression above is the Zeldovich 
approximation to the evolution.  
Also note that, if all three axis are initially the same, then 
all the $\lambda_j$s are the same, and $\lambda = \delta_i/3$.  
In this case, the perturbation is a sphere, and the expression 
above reduces to $A(t) \to A(t_i)\, [a_e(t)/a_i]$, so the 
approximation~(\ref{mowhitesilk}) is exact.  

Figure~\ref{bondmo} compares the full numerical evolution of the 
three axes in this model (solid) with the Zeldovich approximation 
(dotted) and with equation~(\ref{mowhitesilk}) (dashed), when the 
initial values are $(\delta,e,p)=(2/400,0.2,0)$ in an Einstein 
de-Sitter universe.  
This shows clearly that, at late times, equation~(\ref{mowhitesilk}) 
is significantly more accurate than the Zeldovich approximation.  

\begin{figure}[!ht]
\begin{center}
\plotone{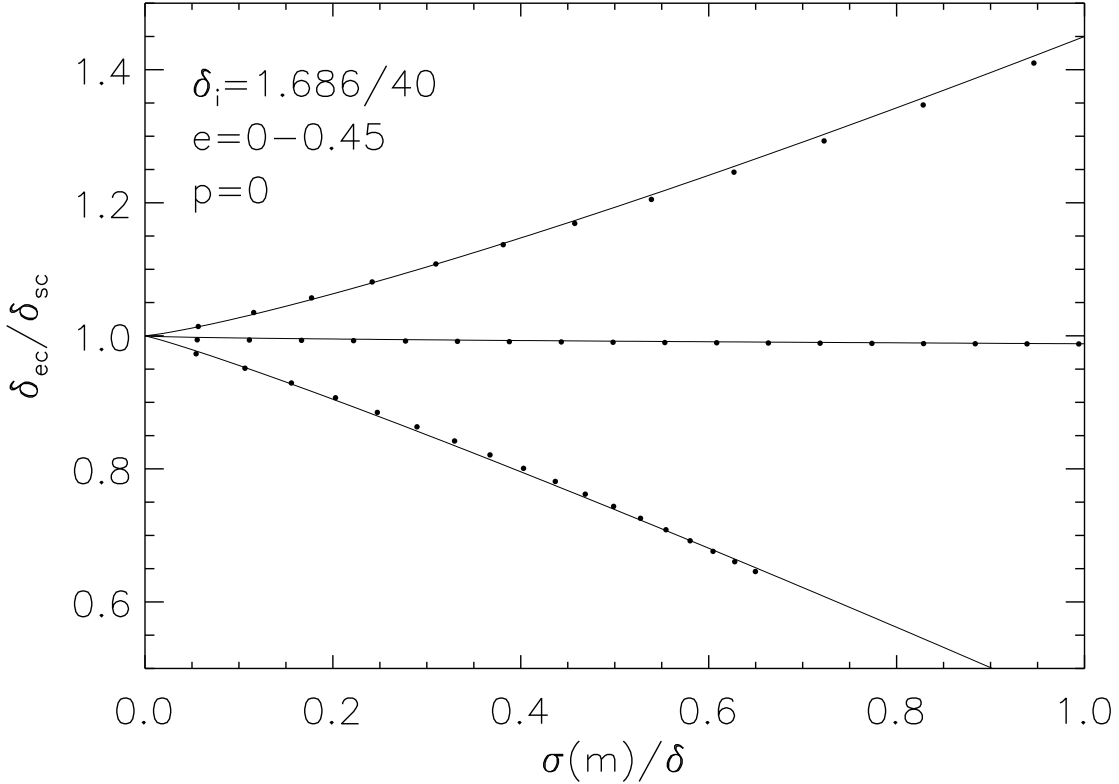}
\end{center}
\caption{Dependence of barrier height on mass for collapse along 
         one, two and three axes (bottom to top).  Symbols show 
         the relations determined from the ellipsoidal collapse 
         model, and curves show the analytic forms given in the 
         main text (equation~ 2). }
\label{shethall}
\end{figure}

The numerical solution allows each axis to shrink to arbitrarily 
small sizes.  In studies of the spherical evolution model it is 
standard to assume that virial equilibrium leads to a non-negligible 
freeze-out radius.  For an Einstein de-Sitter model, this freeze-out 
radius is $A_k = 0.177{a}$, so the final object is 
$18\pi^2\approx 178$ times denser than the background universe.  
\citet{B96} suggested that this same factor could be used in 
the ellipsoidal collapse model as well:  hence, we stop the 
collapse of axis $k$ by hand when $A_k/a=0.17$.  

Thus, in this model, initial values of $\lambda_k$ and $a_i$ yield 
initial values of $\delta_i,e$ and $p$.  The collapse model then 
provides an estimate of the times at which each axis freezes-out.  
In particular, note that if $e$ and $p$ are given, then there is a 
unique value of $\delta_i$ which will produce collapse of the $k$th 
axis at redshift $z$.  
The symbols in Figure~\ref{shethall} show the critical densities 
for collapse at $z=0$ when $p=0$ and $e = (\sigma/\delta)/\sqrt{5}$ 
along one (bottom), two (middle) and three (top) axes.  
In the random walk model discussed in the main text, these values of 
$e$ and $p$, and this relation between the critical density required 
for collapse at $z$ and the initial shape play a central role.  
The solid curves show the simple fits to these relations given in 
equation~( 2).  

Notice that the critical density for collapse of the second axis 
is {\em very} similar to that for a spherical model.  
The approximation~(\ref{mowhitesilk}) provides an easy way to see 
why.  First, note that to second order in $\delta_i$, 
\begin{equation}
 {A_h(t_i)\over a_i} \approx 1 - {\delta_i\over 3} 
                         - 2\,(3e^2 + p^2)\,\left({\delta_i\over 3}\right)^2.
\end{equation}
Second, recall that when $p=0$ then $\lambda_2=\delta_i/3$.  
Third, when $p=0$ and $e = (\sigma/\delta)/\sqrt 5$ (our representative 
values) then the final term in the expression above is $(2/15)\sigma_i^2$, 
so $A_h(t_i) \approx A_2(t_i) - a_i\,(2/15)\sigma_i^2$.  
Therefore, for $\sigma_i\ll 1$, 
\begin{equation}
 {A_2(t)\over A_2(t_i)} \approx {a(t)\over a_i}
\Bigl[1-D(t)\,\lambda_2\Bigr]
- {a(t)A_h(t_i)\over a_i A_2(t_i)} \left[ 1-\frac{D(t) \delta_i}{3} 
                             - \frac{a_e(t)}{{a}(t)} \right]
        \approx {a_e(t)\over a_i};
\end{equation}
in this approximation, the second axis evolves {\em exactly} 
as in a spherical model with initial overdensity $\delta_i$.  

\end{document}